\documentclass[a4paper]{spie}  
\usepackage[utf8]{inputenc}
\usepackage[T1]{fontenc}
\usepackage[english]{babel}
\usepackage[intlimits]{amsmath}
\usepackage{units}
\usepackage{color}
\usepackage{graphicx}
\usepackage{epstopdf}
\usepackage{subfig}
\usepackage{bbm}
\usepackage{multirow}
\usepackage{array}
\usepackage[hyphens]{url}
\usepackage{upgreek}

\newcommand{\Stokes}{\ensuremath{\mathrm{St}}}
\newcommand{\Reynolds}{\ensuremath{\mathrm{Re}}}

\newcommand{\rd}{\ensuremath{\mathrm{d}}}

\newcommand\transpose{^{\mathrm T}}

\newcommand\mat[1]{#1}
\renewcommand{\vec}[1]{\mathbf{#1}}

\title{A microfluidic device for the study of the orientational dynamics of microrods}
\author{Y. N. Mishra\supit{a,b},
J. Einarsson\supit{a},
O. A. John\supit{a},
P. Andersson\supit{a},\\
B. Mehlig\supit{a}, and D. Hanstorp\supit{a}
\skiplinehalf
\supit{a}Department of Physics, University of Gothenburg, 412 96 Gothenburg, Sweden \\
\supit{b}Centre of Excellence in Lasers \& Optoelectronic Sciences, CUSAT, Cochin-682022, India
}

\begin{document}
\maketitle

\begin{abstract}
We describe a microfluidic device for studying the orientational dynamics of microrods.
The device enables us to experimentally investigate the tumbling of microrods immersed
in the shear flow in a microfluidic channel
with a depth of $\unit[400]{\upmu m}$ and a width of $2.5\,$mm. The orientational dynamics was
recorded using a $20$X microscopic objective and a CCD camera. The microrods were produced
by shearing microdroplets of photocurable epoxy resin. We show different
examples of empirically observed tumbling. On the one hand we find
that short stretches of the experimentally determined time series are well described by fits
to solutions of Jeffery's approximate equation of motion [Jeffery, Proc. R. Soc. London. 102 (1922), 161-179]. On the other hand we find that the empirically observed trajectories drift
between different solutions of Jeffery's equation. 
We discuss possible causes of this {\rm orbit drift}.

\end{abstract}



\section{INTRODUCTION}\label{sec:introduction}

In recent years there has been a significant development of lab-on-chip devices designed to be used as biological assays \cite{schasfoort2004} in environmental monitoring for the detection of toxic materials \cite{adams2005} as well as for the assembly of nano- and microscale objects into more complex systems\cite{dittrich2995, Srinivasan2004}.  The main reason
for miniaturisation is to reduce the cost of production for the devices, 
as well as to reduce the volume needed to analyse samples. Last but not least,
the lithographic method used to make the devices allows the production of arbitrary channel structures at a low cost.

There have been numerous investigations where microfludic devices have been used to study the properties of microparticles. For example, Eriksson  \emph{et al.} used a microfluidic system in combination with  optical tweezers to investigate salt stress in yeast cells\cite{eriksson2007}. Alrifaiy \emph{et al.} conducted electrophysiological investigations of individual cells \cite{alrifaiy2010}. Tegenfeldt and coworkers have developed a method to sort cells that differ in both size\cite{long2008} and morphology\cite{holm2011} by means of their motion in an array of obstacles. These studies rely on a precise understanding of the dynamics of particles in microfluidic devices. The translational dynamics of microparticles is often well described by center-of-mass advection.

For non-spherical particles the orientational dynamics must also be considered.
Already in 1922 Jeffery derived an approximate equation of motion for the orientational
dynamics of non-spherical particles\cite{jeffery1922}. Jeffery's solution predicts that elongated axisymmetric particles mostly align with the flow direction, but periodically
rapidly turn around 180 degrees. This motion is referred to as \lq tumbling' 
in the literature.
Jeffery's equations form the foundation for studying the rheology of suspensions of non-spherical particles,
together with a diffusion approximation\cite{hinch1972,brenner1974}. We refer to Petrie\cite{petrie1999} for a contemporary review of fibre-suspension rheology and the applications of Jeffery's equations in this field. Recently, pattern formation
by elongated particles in random flows was investigated using Jeffery's theory\cite{wilkinson2009,bezuglyy2010,wilkinson2011}, with applications to flow visualisation (rheoscopy).
With much past, present and future research based on Jeffery's equations, it is necessary to study how accurate Jeffery's theory is in predicting the orientational dynamics of a single microparticle.

The aim of this investigation is to design a device that makes it possible to study the orientational dynamics of a single non-spherical particle, such that the results can be compared to
theoretical predictions. In this article we describe a microfluidic device allowing us to study single-particle orientational dynamics advected in a channel flow. The channel is produced in PDMS, and the particles are produced using a liquid-liquid dispersion technique. By video microscopy with a motorised stage we observe the orientational motion of particles advected along the channel. A recent experiment\cite{kaya2009} displays short sequences of tumbling motion of \textit{Escherichia coli} bacteria. It is demonstrated that the bacteria follow Jeffery's theory as they tumble a few times.
Our experimental setup enables us to follow a particle over longer distances. We show examples where stretches of the experimentally determined time series are well described by fits
to solutions of Jeffery's equation of motion. But we also demonstrate that the empirically observed trajectories drift
between different solutions of Jeffery's equation. 
We discuss possible causes of this effect: thermal noise, random fluctuations of the flow
due to imperfections in the channel, and the possibility that the particles are not perfectly axisymmetric. 
We argue that inertial effects, due to particle or fluid inertia, are less important.

We see the work described in the following as a step towards a device that makes it possible to investigate the dynamics of a single particle with full parameter control. Our aim in future studies is to control the particle shape, its initial conditions, and the flow conditions. Particles produced with lithographic\cite{guan2005} and photo-polymerisation \cite{kelemen2006} methods
are ideally suited to achieve precisely tailored particles. The initial conditions of a particle, both with respect to position and orientation, can be controlled by means of optical tweezers.
The viscosity of the medium is conveniently varied by using different concentrations of a glycerol/water mixture.

This article is organised as follows. In Sec.~\ref{sec:method}, we describe the experimental setup, the fabrication method for rod-like microparticles and the data analysis procedure. Our measured experimental data is presented in Sec.~\ref{sec:results}. The results are compared to theory and discussed in Sec.~\ref{sec:discussion}. We end the article with concluding remarks and an outlook in
Sec.~\ref{sec:conclusion}.

\section{MATERIALS AND METHODS}\label{sec:method}
\subsection{Microrod fabrication}
\begin{figure}
\begin{center}
\subfloat[A transmission image of polymer microrods imaged through a 20X microscope objective. The distance between minor ticks is 10 $\unit{\upmu m}$. ]{
\includegraphics[width=5.5cm]{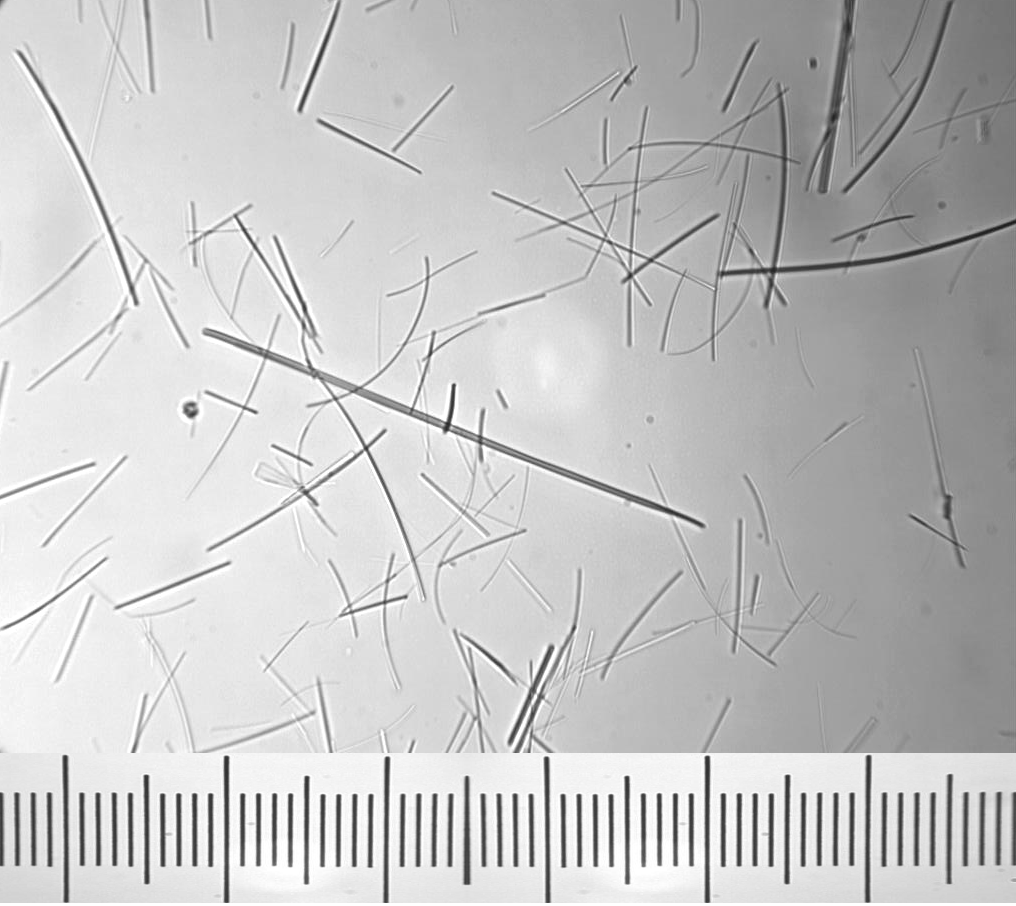}
}\qquad
\subfloat[Measured length distribution of particles. The particles used in this work are approximately 30 $\unit{\upmu m}$ long.]{
\includegraphics[width=6cm]{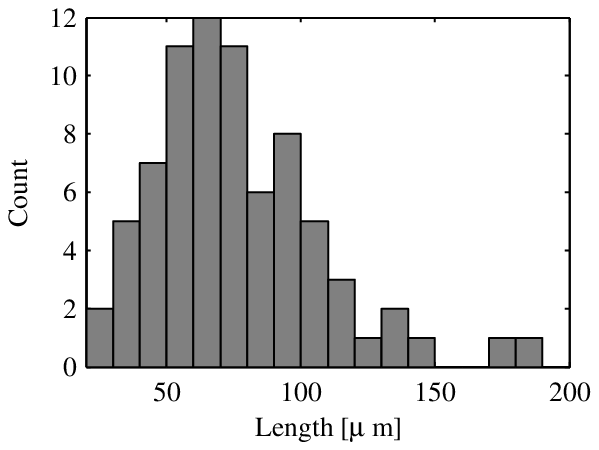}
}
\end{center}
\caption{\label{fig:rod_image} Polymer microrods.}
\end{figure}

The polymer microrods used in this experiment are produced by a liquid-liquid dispersion technique \cite{alargova2004}. A photocurable epoxy resin, SU-8, diluted in solution of $\gamma$-butyrolactone (GBL), is introduced into a 1:1 mixture of ethylene glycol and glycerol while stirring at $\unit[600]{rpm}$. Using a syringe, the SU-8 solution is injected into the spinning liquid where it forms 
microdroplets. These droplets are quickly drawn out to $\unit{\upmu m}$-sized rods by the shear forces in the spinning glycol/glycerole mixture. As the GBL transfers from the SU-8 solution into the
surrounding mixture, the resin solidifies in the shape of microrods. After 10 minutes of stirring at constant rotation the microrod solution is introduced to irradiation of ultraviolet radiation with a wavelength of $\unit[365]{nm}$ for 15 minutes to experience an irreversible cross linking of the photocurable resin. Thereafter, the polymer microrods are separated from
the non-solidified polymer solution using repeated dilution with deionised water and centrifugation. Finally, the microrods are suspended in a solution consisting of a mixture of water and glycerol. The polymer microrods have a tendency to entangle or coalesce in the solution. This problem is solved by adding a few percent of surfactant to the solutions, which reduces
the effect of entanglement and coalescence\cite{alargova2006}. Fig.~\ref{fig:rod_image}a shows a transmission image of the produced microrods using a 20X microscope.
The rods produced using the method described above have a high aspect ratio with a rather wide size distribution. Fig.~\ref{fig:rod_image}b shows the measured size distribution in a sample of microrods.

\subsection{Design and fabrication of microfluidic channels}
A single-inlet/single-outlet  microfluidic device ($\unit[2.5]{mm}$ wide and $\unit[400]{\upmu m}$ deep) is produced in PDMS. A molding form is first created in aluminum using a high precision milling machine. PDMS is molded in this form, cured, removed and finally hole-punched. The PDMS is irreversibly sealed to a plasma treated glass microscope slide with a thickness of $\unit[1]{mm}$. This relatively thick glass is  selected over the standard cover glass in order to avoid mechanical deformation of the cover glass caused by the PDMS. Initial work using a thin cover glass showed that the cover glass deformed. This caused the images of the moving particle to get out of focus as it was tracked along the microchannel.

\subsection{Experimental setup and procedure}
\begin{figure}
\begin{center}
\vspace{1cm}
\includegraphics[width=17cm]{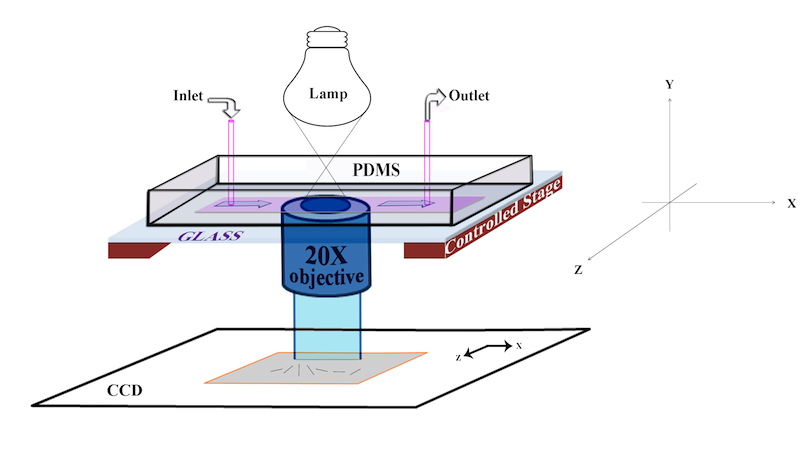}
\end{center}
\caption{\label{fig:experimental_setup} The experimental setup with coordinate system.}
\end{figure}

Our experimental setup is schematically shown in Fig.~\ref{fig:experimental_setup}. Microrods moving in the microfluidic channel are manually tracked using a motorised stage.
The viscosity of the fluid is selected by using a mixture of water and glycerol. A 1:1 ratio is used in all the experiments described below. The water/glycerol mixture is
pumped into the microchannel with a calibrated syringe pump (CMA 400/ Microdialysis) at flows ranging from $\unit[5]{\upmu l/min}$ up to $\unit[11]{\upmu l/min}$. A  20X microscope objective (NA = $0.28$) is used to image and record the microrods' motion onto a CCD camera (Leica DFC 350 FX). The acquired videos are analysed using  MATLAB (Mathworks, Natick, MA, USA).

\subsection{Data analysis}\label{sec:data_analysis}
\begin{figure}[t]
\begin{center}
\begin{tabular}{c}
\includegraphics[width=17cm]{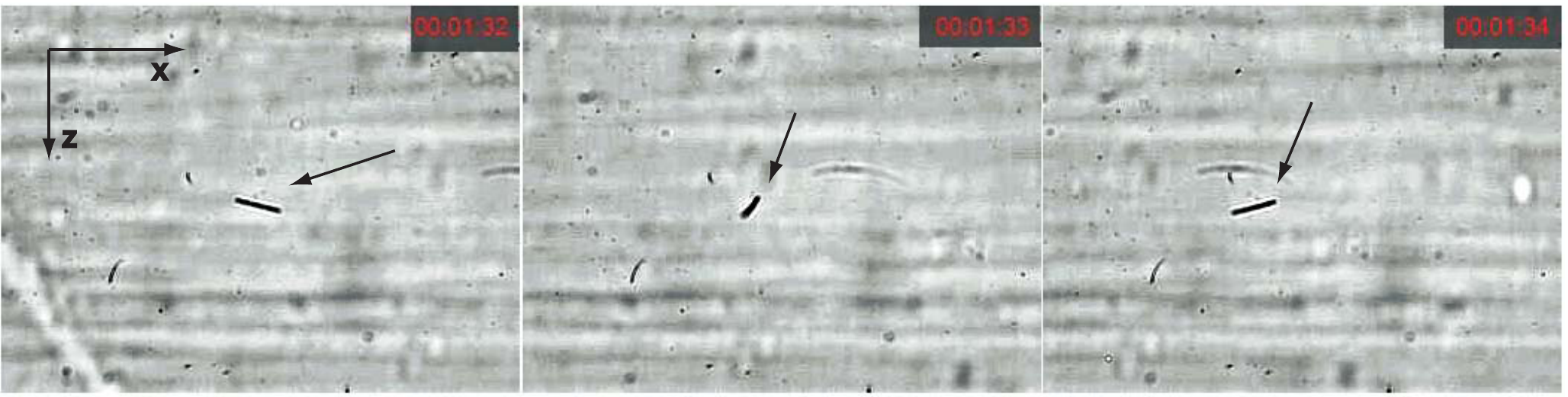}
\end{tabular}
\end{center}
\caption{\label{fig:video_snapshot}Three  frames from video microscopy. The images are in sequence and the time between subsequent frames are approximately one second, corresponding to $100$ frames.
The image  is \unit[$348\times260$]{pixels} wide and high, respectively. Microrods move from left to right, and the camera is following. The particular microrod in focus is indicated by
an arrow in each frame.}
\end{figure}

The camera used in the experiment produces a gray-scale video captured at \unit[100]{Hz} at a resolution of \unit[$348 \times 260$]{pixels}. A typical sequence of one particle during its movement in the channel is several minutes long. Fig.~\ref{fig:video_snapshot} shows three  frames from one experiment presented in this article. The main observable is the orientation of the particle as a function of time. The orientation of an axisymmetric particle can be described by a single vector along the particle's symmetry axis. We define the unit vector $\vec{n}(t)$ as the direction of the particle's symmetry axis at time $t$.

The particle orientation is described in a Cartesian coordinate system where the $x$-axis is along the channel, the $y$-axis pointing down into the channel, into the camera, and the $z$-axis perpendicular to the channel direction in the camera plane. The coordinate system is shown in Fig.~\ref{fig:experimental_setup}. It is important to keep in mind that the video captures a two-dimensional projection of the three-dimensional system. The component of $\vec{n}(t)$ in the $y$-direction (see Fig.~\ref{fig:experimental_setup}), can therefore only be determined in magnitude, but not in sign.

We employ the following procedure to analyse the video data in two steps.
\begin{enumerate}
   \item The endpoints of the particle are identified in each frame.
   \item The endpoints in the camera plane are used to compute the orientation vector $\vec{n}(t)$.
\end{enumerate}
The first step, identification, is performed by assuming that we know the endpoints in the previous video frame. A neighborhood of the particle is cropped from the video and converted to a binary, black-and-white image by Otsu's method \cite{otsu1979}. All remaining connected domains are fitted to ellipses by means of principal-component analysis, and the ellipse closest to the previous frame is chosen. Then the video is advanced one step, whereafter the procedure is repeated. This procedure may fail for several reasons. For example there may be no connected domains in the binary image, or the ellipse closest to the previously found may in fact be the wrong one. On failure the program stops the automatic identification. In this case the endpoints of the rod must be determined manually and entered into the program.

The principal-component analysis cannot distinguish the difference between the two ends of the particle. It only  gives the angle of inclination in the interval $[-90,90]$ degrees. Therefore we manually mark where the direction defined by the endpoints actually pass the $90$ degree inclination. After this step we have acquired a time series of the distinguishable endpoints of the particle.

The particle projection of $\vec{n}(t)$ onto the camera plane and the length of the microrod can then be used to determine the full orientation vector $\vec{n}(t)$, up to the sign in the $y$-component which remains undetermined.

\section{RESULTS} \label{sec:results}
\begin{figure}[t]
\vspace{1cm}
\begin{center}
\includegraphics[width=17cm]{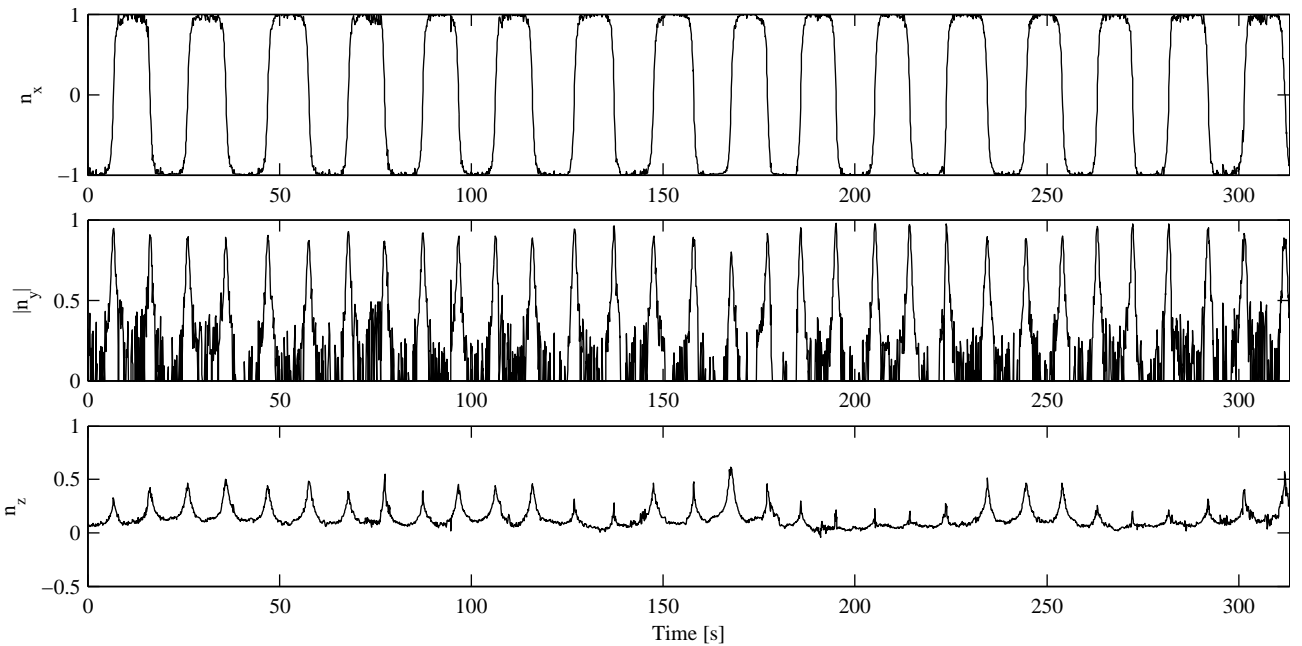}
\end{center}
\caption{\label{fig:example_of_timeseries}Components of the unit vector $\vec{n}(t)$ as a function of time. The top panel shows the $x$-component, which is along the flow in the channel. The middle panel shows the magnitude of $y$-component, which is perpendicular to the camera plane. The bottom panel, finally, shows the $z$-component, or spanwise in the flow. The value of $n_z$ as the microrod
passes through zero yields a measurement of which particular orbit the microrod is following.}
\end{figure}
The experimental data obtained in this work are time series of the unit vector $\vec{n}(t)$ describing the microrod orientation. An example is shown in Fig.~\ref{fig:example_of_timeseries}. The $x$-, $y$- and $z$-components of $\vec{n}(t)$, denoted $n_x$, $n_y$ and $n_z$, are shown separately. The $x$-direction is along the flow and typically stays close to $n_x=\pm 1$ with periodic switches in sign. This type of motion is referred to as tumbling motion, as explained in the introduction. The $y$-component is perpendicular to the camera ($x-z$) plane. It has to be measured by variations in the length of the microrod in the camera plane which presents two problems. First, it only gives information on the magnitude of the $y$-component. Second, the signal is very noisy when the microrod is aligned in the $x-z$-plane, which is the case in a large part of the time series observed.

\section{DISCUSSION}\label{sec:discussion}
In Fig.~\ref{fig:example_of_timeseries} we observe long periods of alignment of the orientation vector $\vec{n}(t)$ along the flow direction, interspersed regularly by tumbling events where the particle quickly turns 180 degrees and aligns in the opposite direction. Between such events we have $|n_x| \approx 1$,
and consequently both $n_y$ and $n_z$ are close to zero during these periods. During the rapid tumbling phases,  typically both $n_y$ and $n_z$ vary. But we have also observed motion purely in $n_y$.

\begin{figure}
\begin{center}
\begin{tabular}{c}
\includegraphics[width=13.5cm]{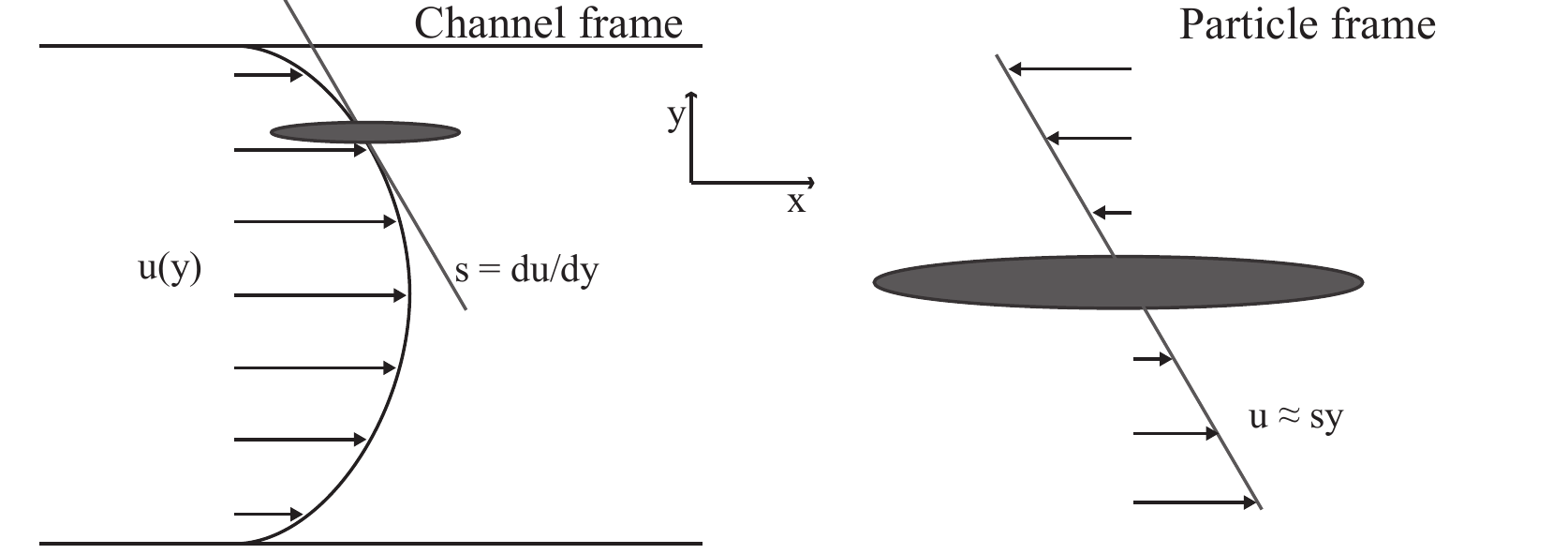}
\end{tabular}
\end{center}
\caption{\label{fig:linearisation} Schematic drawing of the flow profile $u(y)$. Note that the view shown in this figure is perpendicular to the view seen through the microscope objective. The left part of the figure shows the flow profile over the depth of the channel, with the tangent if the profile indicated  at the position of a particle. The right part of the figure  shows a linear approximation of the flow in a coordinate frame translating with the particle along the channel.}
\end{figure}

\begin{figure}
\centering
\subfloat[Jeffery orbits]{
   \includegraphics[width=5cm]{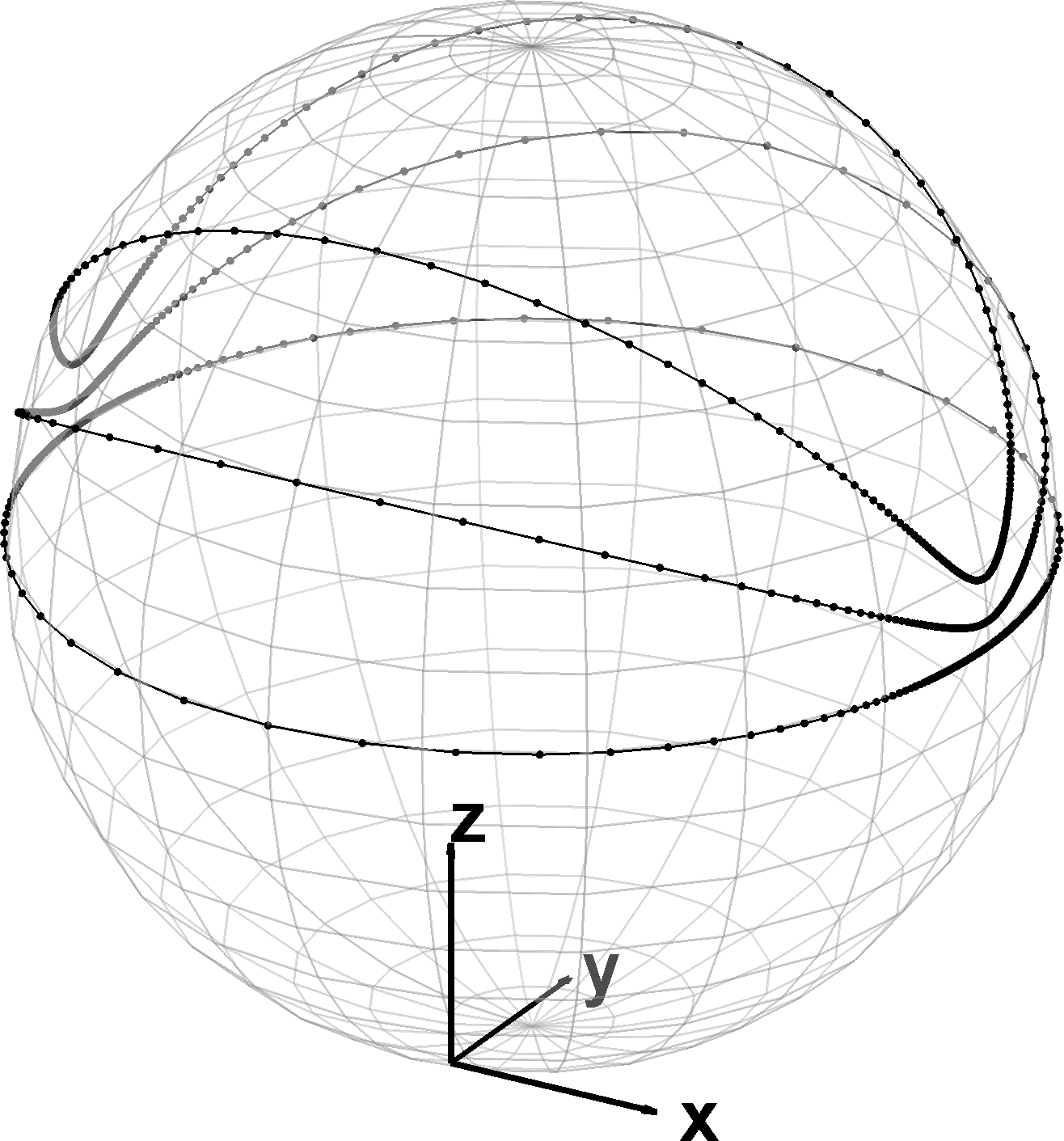}
 }
 \qquad
\subfloat[As seen from camera]{
   \includegraphics[width=5cm]{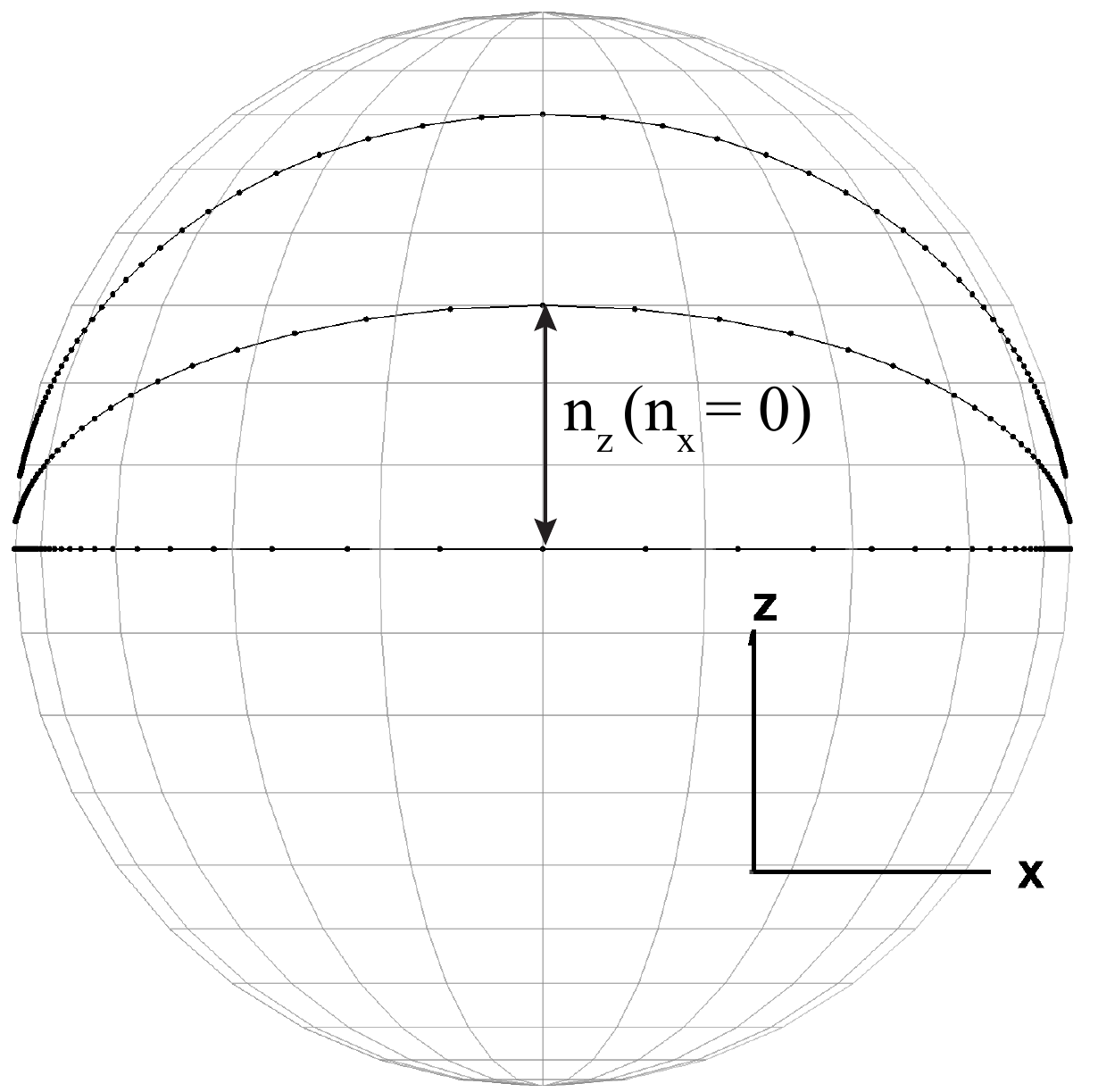}\label{fig:jeffery_example_b}
 }
 \vspace{1em}
\caption{\label{fig:jeffery_example}Visualisation of Jeffery orbits from Eq.~\eqref{eq:explicit_jeffery} with $\lambda=10$ and $\theta_0=\frac{\pi}{2}x_i$, $x_i=0, 0.3, 0.6$ from bottom to top. (a) illustrates the closed paths $\vec{n}(t)$ takes on the sphere where the  dots are plotted for equidistant  times. (b) shows the same but from the viewpoint of the camera in our experiment. A measurement of $n_z$ at the midpoint, where $n_x=0$, uniquely determines the orbit of the particle.}
\end{figure}

The following paragraphs describe the quantitative comparison of our results to Jeffery's theory. Jeffery's results rest on the calculation of the hydrodynamic forces and torques on an ellipsoid, in the limit of a viscous linear flow. We use his results for a special case with three assumptions. First, the ellipsoid is axisymmetric. Second, the particle's inertia is negligible. Third, the flow is a shear flow, as depicted in Fig.~\ref{fig:linearisation}. For this case, Jeffery showed that $\vec{n}(t)$ must describe one of infinitely many closed orbits on the unit sphere. Which particular orbit a particle follows is determined by the initial condition. The complete dynamics is determined by the shear direction, the shear strength $s$, and by the aspect ratio $\lambda$ of the microrod.
Examples of orientational dynamics are shown in Fig.~\ref{fig:jeffery_example}.
When the particles are not perfect ellipsoids, $\lambda$ must be interpreted as an effective aspect ratio (see the discussion in Ref.~\citenum{bretherton1962}).

The channel used in this experiment has an aspect ratio of $1$:$6.25$. We can therefore assume that far away from the side walls, there is no component of the flow gradient in the $z$-direction. Further, the particles are short (order of $\unit[10]{\upmu m}$) in comparison to the channel depth (several $\unit[100]{\upmu m}$). Thus, locally the particles experience a shear flow in the $x$-direction, with shear in the $y$-direction and vorticity in the $z$-direction. Fig.~\ref{fig:linearisation} depicts the linearisation around the particle. In this approximation the flow field $\vec{u}$ is locally described by its gradient
\begin{align}\label{eq:flow_gradient}
	\mat{A} = \nabla \vec{u} &= \begin{pmatrix}
		0 & s & 0 \\
		0 & 0      & 0 \\
		0 & 0      & 0 \\
	\end{pmatrix},
\end{align}
where $s$ is the shear strength. Jeffery's equation for an axisymmetric particle is
\begin{equation}
	\frac{\rd \vec{n}}{\rd t} = \mat{B}\vec{n}-\vec{n}(\vec{n}\transpose \mat{B} \vec{n})\textrm{,}\quad \textrm{with }\mat{B}=\frac{1}{1+\lambda^2}(\lambda^2 A - A\transpose).
\end{equation}
It can be shown\cite{szeri1993,wilkinson2009} that this non-linear equation is solved for any time-independent matrix $\mat{A}$ by
\begin{equation}\label{eq:general_jeffery}
	\vec{q}(t) = \exp(\mat{B}t)\vec{n}(0) \quad \textrm{and } \vec{n}(t) = \frac{\vec{q}(t)}{|\vec{q}(t)|},
\end{equation}
where $\vec{n}(0)$ is the initial orientation at $t=0$. If we assume that $n_x=0$ at $t=0$ and insert the flow gradient \eqref{eq:flow_gradient} into the solution \eqref{eq:general_jeffery}
we arrive at
\begin{align}\label{eq:explicit_jeffery}
\vec{n}(t) &=
\frac{1}{\sqrt{\tan^2\theta_0+ \cos^2\omega t+\lambda^2\sin^2\omega t}}
\begin{pmatrix}
   \lambda\sin\omega t \\
   \cos\omega t\\
   \tan\theta_0
\end{pmatrix}, \quad \omega = \frac{s \lambda }{1+\lambda ^2}.
\end{align}
Here $\theta_0$ is the angle between $\vec{n}(0)$ and the $z$-axis when $n_x=0$, effectively implementing an initial condition and choosing an orbit.
The angle $\theta_0$ is therefore referred to as the orbit constant.\footnote{The common notation for the orbit constant in the literature is $C$ (see for example Eq.~(3b) in Ref.~\citenum{leal1971}) which relates to our notation by $C=\cot \theta_0$. However, we prefer $\theta_0$ as it is a bounded variable on $[-\frac{\pi}{2},\frac{\pi}{2}]$ with a clear physical interpretation as an angle.} Examples of the solutions (\ref{eq:explicit_jeffery}) are shown in Fig.~\ref{fig:jeffery_example}.

In general any point $\vec{n}$ on the unit sphere is associated with a unique orbit. In our particular case the most convenient point to measure which orbit a particle is following is at $n_x=0$,
in other words in the middle of a tumbling event. Reading off the peak height in $n_z$ where $n_x$ passes through zero, as indicated in  Fig.~\ref{fig:jeffery_example_b}, uniquely determines
the orbit the microrod is following.

\begin{figure}
\begin{center}
\begin{tabular}{c}
\includegraphics[width=17cm]{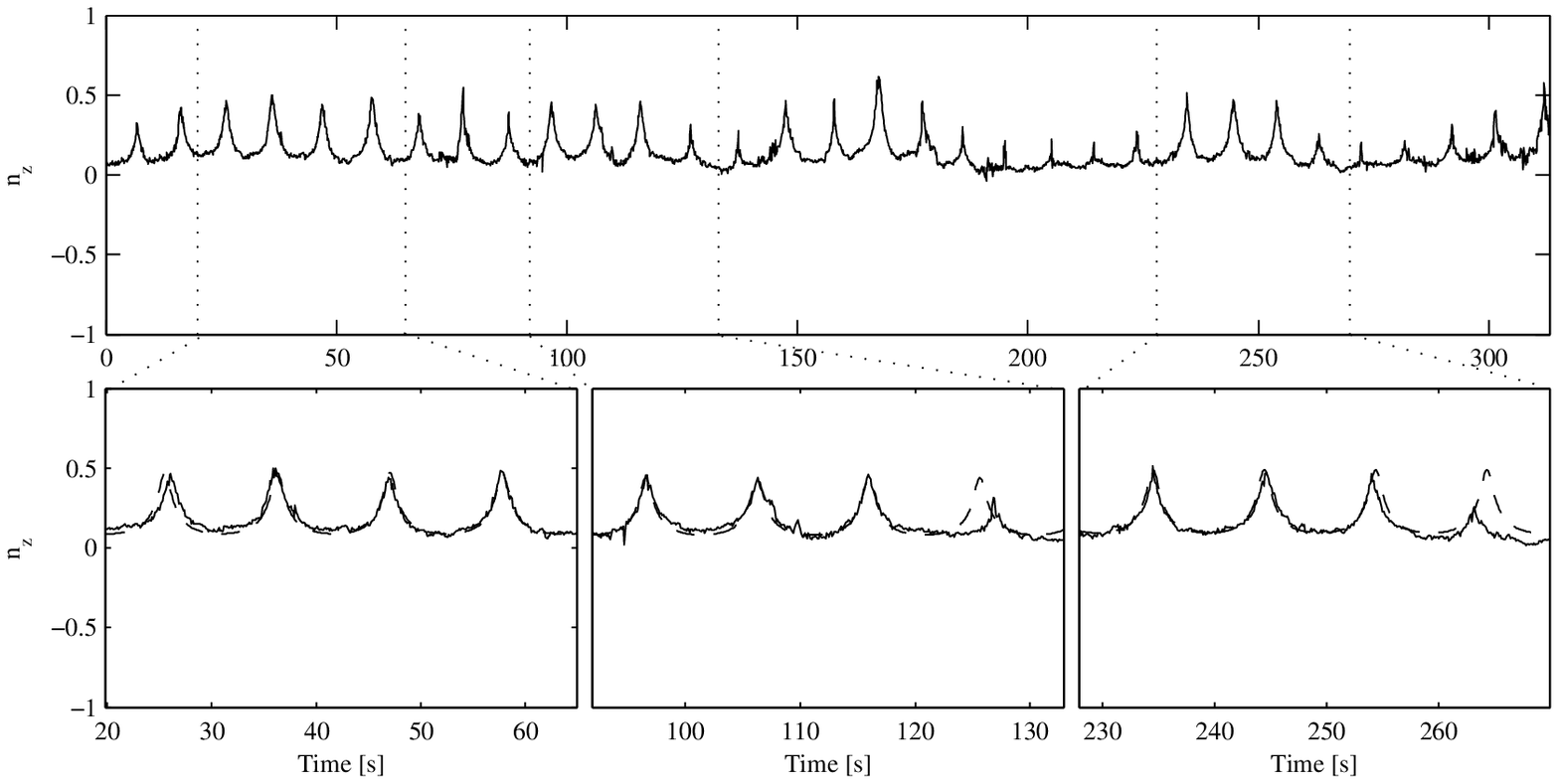}
\end{tabular}
\end{center}
\caption{\label{fig:timeplot1} Sample trajectory showing a relatively stable orbit over time. (top) Component $n_z(t)$ extracted from a video sequence displaying numerous flip events on very similar paths. (bottom) Excerpts from top image, showing in detail the comparison to a suitable Jeffery orbit (dashed) from Eq.~\eqref{eq:explicit_jeffery}. The parameters for the fit are $\lambda=6$ and for the three excerpts $s=1.80, 2.00, 1.95$, respectively, and orbit constant $\theta_0=0.49, 0.46, 0.51$.}
\end{figure}

\begin{figure}
\begin{center}
\begin{tabular}{c}
\includegraphics[width=17cm]{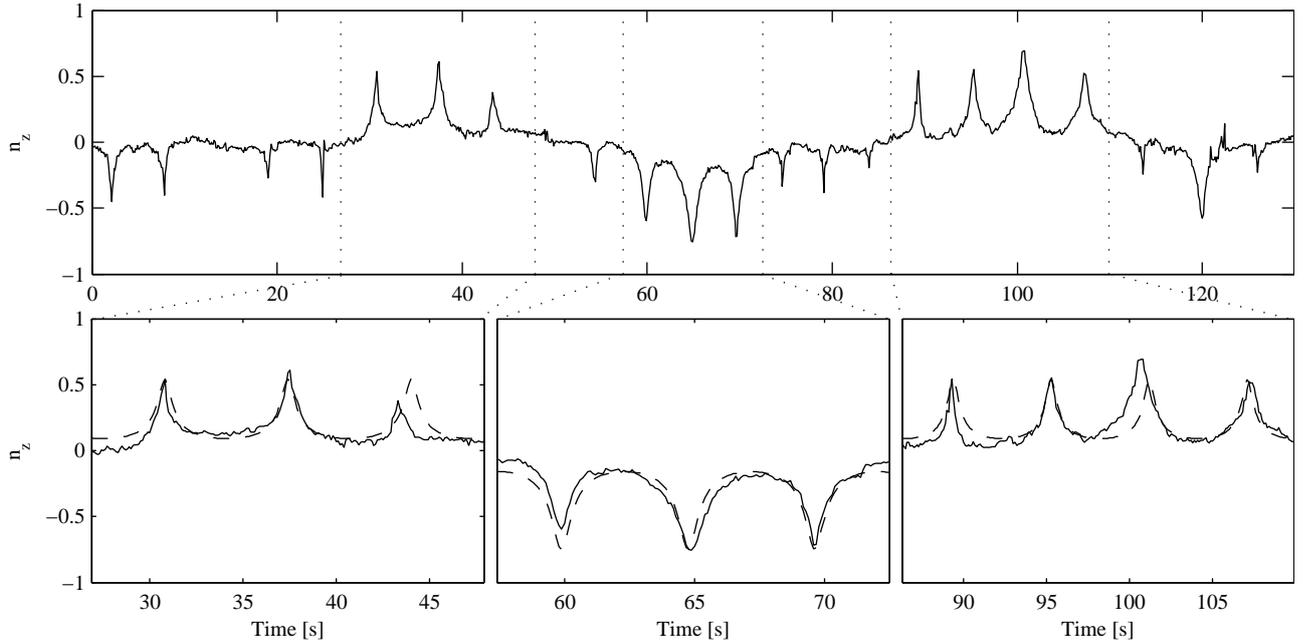}
\end{tabular}
\end{center}
\caption{\label{fig:timeplot2} Sample trajectory showing clear orbit drift. (top) Component $n_z(t)$ extracted from a video sequence displaying numerous flip events with varying paths. (bottom) Excerpts from top image, showing in detail the comparison to a suitable Jeffery orbit (dashed) from Eq.~\eqref{eq:explicit_jeffery}. The parameters for the fit are $\lambda=7$ and for the three excerpts $s=3.40, 4.60, 3.80$, respectively, and orbit constant $\theta_0=0.57, -0.85, 0.57$. }
\end{figure}

Using the $z$-component of Eq.~\eqref{eq:explicit_jeffery} we find segments in our data closely following a Jeffery orbit. We show time series of $n_z$ with such excerpts in Figs.~\ref{fig:timeplot1} and \ref{fig:timeplot2}. Fig.~\ref{fig:timeplot1} depicts a sample where the particle stays on a similar trajectory for several consecutive tumbling events. The sample in Fig.~\ref{fig:timeplot2},
by contrast, shows deviations which we interpret as orbit drift between subsequent tumbling events, as further discussed below.

There are three free parameters to consider when Eq.~\eqref{eq:explicit_jeffery} is used to describe the motion of the particle. They are the shear strength $s$, the effective aspect
ratio $\lambda$, and the orbit constant $\theta_0$. The following paragraphs discuss these parameters in order.

The shear strength $s$ affects the period time of tumbling linearly. The shear strength $s$ is determined by the flow profile in the channel,
and by the position of the microrod in the $y$-direction. Moreover, the shear rate can fluctuate if the syringe pump does not provide perfectly steady flow. Currently
temporal variations of the shear rate are not available with good enough precision, and $s$ must remain a free parameter.
As shown in Figs.~\ref{fig:timeplot1} and \ref{fig:timeplot2}, the estimated $s$ is of order unity. This is consistent with the setup of our experiment where the depth of the channel is $L\approx \unit[400]{\upmu m}$, making the typical flow velocity $v \approx sL/2 = \unit[200]{\upmu m/s}$. The channel is around \unit[5]{cm} long, and it takes the particles around 5 minutes to travel its entire length. A speed of $\unit[1]{cm/minute}$ gives approximately $\unit[150]{\upmu m/s}$. The  observed speed is hence consistent with the estimated shear strength.

The effective aspect ratio $\lambda$ is a dimensionless number describing the shape of an axisymmetric particle. An aspect ratio of $\lambda=1$ corresponds to a sphere, and $\lambda=10$ corresponds to a cigar-shaped particle. For an ellipsoid, $\lambda$ is the actual aspect ratio. For a microrod we expect $\lambda$ to be close to its length divided by its width. The estimates in Figs.~\ref{fig:timeplot1} and \ref{fig:timeplot2} are $\lambda=6$ and $7$, respectively. For both examples presented we measure $\lambda\approx 10$ (see Fig.~\ref{fig:video_snapshot}), in reasonable agreement.
The aspect ratio is a particle property and as such it cannot be time dependent.

The orbit constant $\theta_0$ is a free parameter, a constant of integration, in the derivation of Eq.~\eqref{eq:explicit_jeffery}. The theory gives no preference to any particular orbit, but asserts that $\theta_0$ must be constant in time. In the figures, and as can be seen in Eq.~\eqref{eq:explicit_jeffery}, $\theta_0$ determines the peak height of $n_z$ as discussed above.
Clearly the observed values of $\theta_0$ are not perfectly independent of time. In the context of Jeffery's theory this may be described as drift between different solutions of Eq.~\eqref{eq:explicit_jeffery}, or {\em orbit drift}.
We infer that there must either be effects beyond the purely hydrodynamic forces considered by Jeffery
(possible effects may be inertial, due to either particle inertia -- finite Stokes number -- or fluid inertia 
-- finite Reynolds number). Another reason may be that the particles are not perfectly axisymmetric.
Last but not least the particle dynamics is certainly affected by noise, either
thermal noise or due to flow irregularities. These possibilities are discussed in the following paragraphs.

A fluid Reynolds number based on the channel dimension (order of $\unit[10^2]{\upmu m}$) and the flow speed (order of $\unit[10^2]{\upmu m/s}$) is on the 
order of $\Reynolds=10^{-2}$, for water at room temperature with kinematic viscosity of $\unit[10^{-6}]{m^2/s}$. In our experiments the viscosity is even higher  since we mix glycerol into the liquid. The particles used in this work are neutrally buoyant. The Re-dependence has been studied numerically by several authors\cite{qi2003,yu2007}. Their results indicate that the Jeffery solution based on creeping flow (Re$=0$) is valid well up to $\Reynolds=10^{-1}$ and possibly higher. Lundell and Carlsson\cite{lundell2010} studied numerically the effect of increasing $\Stokes$ at low $\Reynolds$. They found that the particles follow the Jeffery dynamics for moderately small $\Stokes$  with only a very slow orbit drift towards a neutral center orbit (corresponding to $\theta_0=0$ in our notation). On the contrary, we observe that orbit drift can occur quite rapidly, and in any direction, not towards a particular equilibrium. Hence, previous investigations of inertial effects do not provide an
explanation for our empirical results.

The effects of noise on the single-particle orientational dynamics have not received much attention in the past. However, recent years have seen an increased interest in single-particle dynamics, sparked by single-polymer experiments such as those by Smith \emph{et al.}\cite{smith1999} and Schroeder \emph{et al.}\cite{schroeder2005}, followed by theoretical investigations by Chertkov \emph{et al.}\cite{chertkov2005} and Turitsyn\cite{turitsyn2007}. These studies focus on the limit of extremely slender particles\footnote{Actually, the authors investigate flexible polymers, which do not have the same dependence on the aspect ratio. The equations, however, are equivalent to the limit of rigid rods of infinite aspect ratio.}, $\lambda \to \infty$, and assume that there is a random component of the flow gradient, $A$, modeled by a diffusion coefficient. As it turns out the finite aspect ratio $\lambda$ is crucial to the Jeffery dynamics of periodic tumbling. Since our observations suggest that the tumbling motion is indeed periodic, the limit $\lambda \to \infty$ is irrelevant to our experiment.

There have been several studies of angular statistics of large ensembles of particles\cite{hinch1972,brenner1974}. However, these studies focus mainly on the 
rheology of macroscopic suspensions. We can use the expression in Ref.~\citenum{brenner1974} to approximate the rotational diffusion constant of a spheroid, assuming room temperature and a $\unit[10\times 1\times 1]{\upmu m}$ particle in water. We find that $D_r\approx \unit[10^{-2}]{radians/s}$. The rotational diffusion rate should be compared to the typical rate of rotation due to the shear. While the particle is nearly aligned with the flow direction the deterministic rate of rotation is of order $s/(1+\lambda^2)$. With $s\approx 1$ and $\lambda\approx 10$ we find that the deterministic and noise-induced rates of rotation are comparable in magnitude. This discussion leads to the conclusion that noise is an important mechanism to describe our observations.

Finally, we consider the possibility that the particles are not perfectly axisymmetric. Even small deviations from axisymmetry can result in relatively large changes in a particle's rotation\cite{hinch1979,yarin1997}. 
In particular the results of Refs.~\citenum{hinch1979} and \citenum{yarin1997} indicate that weak asymmetry can result in drift between different observed values of the orbit constant. 
In order to experimentally distinguish between the effects of noise and particle asymmetry, it is necessary
to precisely control the particle shape. Furthermore, observing more consecutive tumblings will facilitate
to distinguish periodic modulations due to asymmetry from random drift.

\section{CONCLUSION AND OUTLOOK}\label{sec:conclusion}

To conclude, our results suggest that a plausible way to understand the empirically observed orientational dynamics of single particles may begin in the hydrodynamic theory by Jeffery. We have argued that two additional mechanisms may affect the dynamics significantly. First, noise causes random drift between different Jeffery orbits. This noise could be modeled as a random contribution to the flow-gradient matrix $\mat{A}$. 
The fundamental parameters to explore in this case are the aspect ratio $\lambda$, and the Pecl\'et number. The Pecl\'et number, as is well known, is a dimensionless ratio between the magnitudes of the deterministic and random components of the flow. The second mechanism that deserves further investigation is that weak asymmetry of particles can induce drift in the orbit constant. The case of non-axisymmetric particles is part of the Jeffery theory, of which Eq.~\eqref{eq:general_jeffery} is a special case.

Future experiments will focus on an increased control of the relevant parameters. Work in progress includes coupling a piezoelectric element to the glass cover, which can be used to induce ultrasonic noise in the liquid. With particle velocimetry and volumetric flow measurements we can determine the shear strength. The initial position and orientation of the microrod can be controlled by installing optical tweezers in the setup. In parallel we are developing a photopolymerisation technique to consistently produce high precision particles of arbitrary shape. As we have argued, precise control over the particle shape is crucial to distinguish the effects of particle shape from random orbit drift. 
\acknowledgements
Financial support from the Swedish Research Council, 
the G\"oran Gustafsson Foundation for Research in Natural Sciences and Medicine, and the platform ``Nano-particles in an interactive environment''
is gratefully acknowledged. 


\bibliography{report}   

\begin{thebibliography}{10}

\bibitem{schasfoort2004}
Schasfoort, R., ``{Proteomics-on-a-chip: the challenge to couple lab-on-a-chip
  unit operations},'' {\em {Expert review of proteomics}}~{\bf {1}}({1}),
  {123--132} ({2004}).

\bibitem{adams2005}
Adams, M., Loncar, M., Scherer, A., and Qiu, Y., ``{Microfluidic integration of
  porous photonic crystal nanolasers for chemical sensing},'' {\em {IEEE
  Journal on Selected Areas in Communications}}~{\bf {23}}({7}),  {1348--1354}
  ({2005}).

\bibitem{dittrich2995}
Dittrich, P., Jahnz, M., and Schwille, P., ``{A new embedded process for
  compartmentalized cell-free protein expression and on-line detection in
  microfluidic devices},'' {\em {Chembiochem}}~{\bf {6}}({5}),  {811+}
  ({2005}).

\bibitem{Srinivasan2004}
Srinivasan, V., Pamula, V., and Fair, R., ``{An integrated digital microfluidic
  lab-on-a-chip for clinical diagnostics on human physiological fluids},'' {\em
  {Lab on a chip}}~{\bf {4}}({4}),  {310--315} ({2004}).

\bibitem{eriksson2007}
Eriksson, E., Enger, J., Nordlander, B., Erjavec, N., Ramser, K., Goksor, M.,
  Hohmann, S., Nystrom, T., and Hanstorp, D., ``{A microfluidic system in
  combination with optical tweezers for analyzing rapid and reversible
  cytological alterations in single cells upon environmental changes},'' {\em
  {Lab on a Chip}}~{\bf {7}}({1}),  {71--76} ({2007}).

\bibitem{alrifaiy2010}
Alrifaiy, A., Bitaraf, N., Lindahl, O., and Ramser, K., ``{Development of
  Microfluidic System and Optical Tweezers for electrophysiological
  investigations of an individual cell},'' in [{\em {Optical trapping and
  optical micromanipulation VII}}{\nolinebreak\hspace{0.1em}]},  {Dholakia, K
  and Spalding, GC}, ed., {\em {Proceedings of SPIE-The International Society
  for Optical Engineering}} {\bf {7762}} ({2010}).

\bibitem{long2008}
Long, B.~R., Heller, M., Beech, J.~P., Linke, H., Bruus, H., and Tegenfeldt,
  J.~O., ``{Multidirectional sorting modes in deterministic lateral
  displacement devices},'' {\em {Physical Review E}}~{\bf {78}} ({2008}).

\bibitem{holm2011}
Holm, S.~H., Beech, J.~P., Barrett, M.~P., and Tegenfeldt, J.~O., ``{Separation
  of parasites from human blood using deterministic lateral displacement},''
  {\em {Lab on a Chip}}~{\bf {11}}({7}),  {1326--1332} ({2011}).

\bibitem{jeffery1922}
Jeffery, G.~B., ``The motion of ellipsoidal particles immersed in a viscous
  fluid,'' {\em Proceedings of the Royal Society of London. Series A}~{\bf
  102}(715),  161--179 (1922).

\bibitem{hinch1972}
Hinch, E.~J. and Leal, L.~G., ``The effect of brownian motion on the
  rheological properties of a suspension of non-spherical particles,'' {\em
  Journal of Fluid Mechanics}~{\bf 52}(04),  683--712 (1972).

\bibitem{brenner1974}
Brenner, H., ``Rheology of a dilute suspension of axisymmetric brownian
  particles,'' {\em International Journal of Multiphase Flow}~{\bf 1}(2),
  195--341 (1974).

\bibitem{petrie1999}
Petrie, C., ``The rheology of fibre suspensions,'' {\em Journal of
  Non-Newtonian Fluid Mechanics}~{\bf 87}(2-3),  369 -- 402 (1999).

\bibitem{wilkinson2009}
Wilkinson, M., Bezuglyy, V., and Mehlig, B., ``Fingerprints of random flows?,''
  {\em Physics of Fluids}~{\bf 21}(4),  043304 (2009).

\bibitem{bezuglyy2010}
Bezuglyy, V., Mehlig, B., and Wilkinson, M., ``Poincaré indices of rheoscopic
  visualisations,'' {\em Europhysics Letters}~{\bf 89}(3),  34003 (2010).

\bibitem{wilkinson2011}
Wilkinson, M., Bezuglyy, V., and Mehlig, B., ``Emergent order in rheoscopic
  swirls,'' {\em Journal of Fluid Mechanics}~{\bf 667},  158--187 (2011).

\bibitem{kaya2009}
Kaya, T. and Koser, H., ``Characterization of hydrodynamic surface interactions
  of \textit{Escherichia coli} cell bodies in shear flow,'' {\em Phys. Rev.
  Lett.}~{\bf 103},  138103 (2009).

\bibitem{guan2005}
Guan, J., Chakrapani, A., and Hansford, D., ``{Polymer microparticles
  fabricated by soft lithography},'' {\em {Chemistry of materials}}~{\bf
  {17}}({25}),  {6227--6229} ({2005}).

\bibitem{kelemen2006}
Kelemen, L., Valkai, S., and Ormos, P., ``{Integrated optical motor},'' {\em
  {Applied Optics}}~{\bf {45}}({12}),  {2777--2780} ({2006}).

\bibitem{alargova2004}
Alargova, R., Bhatt, K., Paunov, V., and Velev, O., ``{Scalable synthesis of a
  new class of polymer microrods by a liquid - Liquid dispersion technique},''
  {\em {Advanced Materials}}~{\bf {16}}({18}),  {1653} ({2004}).

\bibitem{alargova2006}
Alargova, R., Paunov, V., and Velev, O., ``{Formation of polymer microrods in
  shear flow by emulsification - Solvent attrition mechanism},'' {\em
  {Langmuir}}~{\bf {22}}({2}),  {765--774} ({2006}).

\bibitem{otsu1979}
Otsu, N., ``A threshold selection method from gray-level histograms,'' {\em
  IEEE Transactions on Systems, Man and Cybernetics}~{\bf 9}(1),  62--66
  (1979).

\bibitem{bretherton1962}
Bretherton, F.~P., ``The motion of rigid particles in a shear flow at low
  reynolds number,'' {\em Journal of Fluid Mechanics}~{\bf 14}(02),  284--304
  (1962).

\bibitem{szeri1993}
Szeri, A.~J., ``Pattern formation in recirculating flows of suspensions of
  orientable particles,'' {\em Philosophical Transactions: Physical Sciences
  and Engineering}~{\bf 345}(1677),  pp. 477--506 (1993).

\bibitem{leal1971}
Leal, L.~G. and Hinch, E.~J., ``The effect of weak brownian rotations on
  particles in shear flow,'' {\em Journal of Fluid Mechanics}~{\bf 46}(04),
  685--703 (1971).

\bibitem{qi2003}
Qi, D. and Luo, L., ``Rotational and orientational behaviour of
  three-dimensional spheroidal particles in couette flows,'' {\em Journal of
  Fluid Mechanics}~{\bf 477},  201--213 (2003).

\bibitem{yu2007}
Yu, Z., Phan-Thien, N., and Tanner, R.~I., ``Rotation of a spheroid in a
  couette flow at moderate reynolds numbers,'' {\em Phys. Rev. E}~{\bf 76},
  026310 (2007).

\bibitem{lundell2010}
Lundell, F. and Carlsson, A., ``Heavy ellipsoids in creeping shear flow:
  Transitions of the particle rotation rate and orbit shape,'' {\em Phys. Rev.
  E}~{\bf 81},  016323 (2010).

\bibitem{smith1999}
Smith, D.~E., Babcock, H.~P., and Chu, S., ``Single-polymer dynamics in steady
  shear flow,'' {\em Science}~{\bf 283}(5408),  1724--1727 (1999).

\bibitem{schroeder2005}
Schroeder, C.~M., Teixeira, R.~E., Shaqfeh, E. S.~G., and Chu, S.,
  ``Characteristic periodic motion of polymers in shear flow,'' {\em Phys. Rev.
  Lett.}~{\bf 95},  018301 (2005).

\bibitem{chertkov2005}
Chertkov, M., Kolokolov, I., Lebedev, V., and Turitsyn, K., ``Polymer
  statistics in a random flow with mean shear,'' {\em Journal of Fluid
  Mechanics}~{\bf 531},  251--260 (2005).

\bibitem{turitsyn2007}
Turitsyn, K., ``Polymer dynamics in chaotic flows with a strong shear
  component,'' {\em Journal of Experimental and Theoretical Physics}~{\bf 105},
   655--664 (2007).
\newblock 10.1134/S1063776107090245.

\bibitem{hinch1979}
Hinch, E.~J. and Leal, L.~G., ``Rotation of small non-axisymmetric particles in
  a simple shear flow,'' {\em Journal of Fluid Mechanics}~{\bf 92}(03),
  591--607 (1979).

\bibitem{yarin1997}
Yarin, A., Gottlieb, O., and Roisman, I., ``Chaotic rotation of triaxial
  ellipsoids in simple shear flow,'' {\em Journal of Fluid Mechanics}~{\bf
  340},  83--100 (1997).

\end{thebibliography}
\bibliographystyle{spiebib}   

\end{document}